\documentclass[%
 reprint,
superscriptaddress,
nofootinbib,
 amsmath,amssymb,
 aps,
floatfix,
prd,
]{revtex4-2}
\usepackage{graphicx}
\usepackage{dcolumn}
\usepackage{bm}
\usepackage{float}
\usepackage[usenames,dvipsnames]{color}

\begin{document}
\newcommand{\bea}{\begin{eqnarray}}
\newcommand{\eea}{\end{eqnarray}}
\newcommand{\beq}{\begin{equation}}
\newcommand{\eeq}{\end{equation}}
\def\leq{\raise 0.4ex\hbox{$<$}\kern -0.8em\lower 0.62ex\hbox{$-$}}
\def\geq{\raise 0.4ex\hbox{$>$}\kern -0.7em\lower 0.62ex\hbox{$-$}}
\def\lsim{\raise 0.4ex\hbox{$<$}\kern -0.75em\lower 0.65ex\hbox{$\sim$}}
\def\gsim{\raise 0.4ex\hbox{$>$}\kern -0.75em\lower 0.65ex\hbox{$\sim$}}
\def\pm{\,\raise 0.4ex\hbox{$+$}\kern -0.75em\lower 0.65ex\hbox{$-$}\,}

\preprint{APS/123-QED}

\title{Testing the Boundary-to-Bound Correspondence with Numerical Relativity}

\author{Anuj Kankani}
\email{anuj.kankani@mail.wvu.edu}
\affiliation{Department of Physics and Astronomy, West Virginia University, Morgantown, WV 26506, USA
}
\affiliation{Center for Gravitational Waves and Cosmology, West Virginia University, Chestnut Ridge Research Building, Morgantown, WV 26505,
USA
}
\author{Sean T. McWilliams}
\affiliation{Department of Physics and Astronomy, West Virginia University, Morgantown, WV 26506, USA
}
\affiliation{Center for Gravitational Waves and Cosmology, West Virginia University, Chestnut Ridge Research Building, Morgantown, WV 26505,
USA
}

\date{\today}

\begin{abstract}
The Boundary-to-Bound (B2B) correspondence, which connects orbital and radiative observables between bound and unbound orbits, has recently been introduced and demonstrated in the perturbative regime. We produce a large number of numerical relativity simulations of bound and unbound encounters between two nonspinning equal mass black holes in order to test this correspondence in the non-perturbative regime. We focus on testing the radiated energy and angular momentum, as well as orbital parameters such as the period and periastron advance. We find that, across a wide range of eccentricities, the B2B relationships do not hold in the non-perturbative regime, thereby placing a clear limit on the applicability of these relationships. We also approximate the separatrix between bound and unbound relativistic encounters as a function of their initial energies and angular momenta.
\end{abstract}

\maketitle


\section{Introduction}

The full exploitation of the rich astrophysical information carried by gravitational waves requires the generation of accurate gravitational waveforms that can be used to study observed signals. For quasicircular sources, early inspiral orbits can be calculated perturbatively using post-Newtonian (PN) \cite{PN_for_GW_review,PN_2body_review,PN_binary_review,PN_eft_review} or Effective One-Body (EOB) methods \cite{EOB_original_paper,EOB_cite1,EOB_cite2,EOB_cite3,EOB_cite4,EOB_cite5}, while numerical relativity (NR) provides an effective method for accurately simulating the late inspiral, merger and ringdown of two black holes. However, for systems with larger eccentricities, a single orbit might require NR to accurately model the periastron passage, but might also have a long orbital timescale that would make NR too computationally expensive to be practical. Also, the large number of waveforms needed to accurately study the entire parameter space means that NR is too computationally expensive to be used on its own. Semi-analytical approximations like the EOB family of approximants \cite{SEOBNR_cite1,SEOBNR_cite2,SEOBNR_cite3,SEOBNR_cite4} address this issue by using perturbative expansions in the weak field regions and calibrating to NR simulations to better approximate the strong-field interactions. However, simulating only part of a bound orbit in NR presents technical challenges due to the influence of initial data and the difficulty of characterizing orbital parameters.

Until now, only sources with low or no eccentricity have been observed, so the primary focus of gravitational-wave modelers has been to approximate that region of parameter space as accurately as possible.  However, future upgrades to ground-based interferometers and planned space-based observatories like LISA \cite{LISA,LISA_cite2,LISA_cite3} will make it possible to observe gravitational-wave signals from highly eccentric or even hyperbolic systems \cite{lisa_waveform_whitepaper,hyp_search_cite1,hyp_search_cite2,hyp_search_cite3,hyp_search_cite4,hyp_search_cite5,hyp_search_cite6,lisa_astrophysics_whitepaper}  Therefore, future gravitational waveform models must be able to accurately approximate systems with arbitrary eccentricities, so modelers must overcome the challenges of approximating bound relativistic encounters with large eccentricities and long orbital periods. Analytical work on hyperbolic encounters often employs the post-Minkowskian (PM) approximation \cite{PM_cite1,PM_cite2,PM_cite3,PM_cite4,PM_cite5}, which does not require the small velocity assumption used in the PN approximation. A variety of approaches are used to obtain contributions to the PM expansions such as scattering amplitudes \cite{scattering_amplitude_cite1,scattering_amplitude_cite2,scattering_amplitude_cite3,scattering_amplitude_cite4,scattering_amplitude_cite5,4PM_paper3}, effective field theory \cite{PM_eft_cite1,PM_eft_cite2,PM_eft_cite3,4PM_paper2,PM_eft_cite5,4PM_paper6}, and a worldline field theory approach \cite{PM_worldline_cite1,PM_worldline_cite2,PM_worldline_cite3,PM_worldline_cite4}. While there has been significant numerical work done on bound black hole orbits, much less work has been done on unbound orbits. In \cite{patrick_paper,induced_spins_paper2}, the authors studied the spin up of black holes caused by the interaction of two black holes on unbound orbits. The authors of \cite{damour_hopper_scattering,Hopper_damour_spinning_scattering,strong_field_scattering,strong_field_spinning_scattering,Khalil_2022} primarily focused on comparing numerically obtained scattering angles and potentials to analytical estimates.
 
Recently, a series of papers \cite{B2B_paper1,B2B_paper2,B2B_paper3} introduced the ``Boundary-to-Bound'' correspondence (B2B) to connect observables between bound and unbound orbits in the perturbative regime, including most of the available conservative contributions and leading-order radiative effects. These results provide a compelling new avenue to study bound orbits; since hyperbolic encounters in NR are less affected by initial data transients and easier to characterize in terms of orbital parameters than single periastron passages of a bound system, B2B raises the prospect of using relatively inexpensive NR simulations of scattering encounters to provide strong-field information for the relativistic portion of highly eccentric bound systems. Given the computational expense of simulating highly eccentric bound systems for entire orbits, most of which occur in the weak-field regime, B2B could potentially provide a more efficient means for using NR to inform waveform models.

However, as noted by the B2B authors, limitations to the current approach exist, in particular the absence of higher-order radiative corrections and differences in the behavior of nonlinear tail effects on either side of the correspondence, which may limit the applicability of the B2B relationships in full General Relativity (GR). Nevertheless, as was shown in \cite{B2B_paper3}, the incorporation of a large eccentricity limit allows for the inclusion of additional non-local-in-time contributions, raising the possibility of the B2B relationships being applicable for orbits with non-negligible eccentricities.

Furthermore, in \cite{strong_field_scattering,strong_field_spinning_scattering} the authors used NR results to extract the EOB radial potential using the methods of \cite{B2B_paper1}, and found that the result closely matched the 4PM radial potential that incorporated radiative effects. While they only studied the behavior of unbound systems, their work suggested the possibility that not only conservative contributions, but also dissipative ones, might possibly map similarly from the unbound to the bound regions of parameter space. 

In this paper we test some of the key observables in the B2B dictionary for the case of two equal mass nonspinning black holes. In particular, we test whether NR data for unbound systems can be used to obtain both radiative quantities and orbital characteristics for bound binaries.

\subsection{Notation}
Throughout this work, we use $E$ and $J$ to refer to the ADM energy and angular momentum of the system. Given the individual masses of the two black holes, $m_1$ and $m_2$, the total mass, reduced mass and the symmetric mass ratio are defined as 
\bea
M &=& m_1 + m_2\,,\\
\mu &=& \frac{m_1 m_2}{m_1 + m_2}\,,\\
\nu &=& \frac{m_1 m_2}{(m_1 + m_2)^2}\,.
\eea
The reduced binding energy and angular momentum are defined as 
\bea
    \varepsilon &=& \frac{E-M}{\nu M}\,,\\
    j &=& \frac{J}{GM\mu}\,.
\eea
Finally, we note that the Newtonian eccentricity associated with a conservative system, 
\beq
e_\mathrm{N}^2 = 1 + 2\varepsilon j^2
\eeq
is distinct from the eccentricity $e$ that we use to characterize our fully relativistic bound orbits.

While we test both conservative and dissipative contributions to our observables, unless explicitly written, our analytical information will include only the conservative contributions. Therefore, if we refer to the fourth post-Minkowskian order (4PM) scattering angle, we are referring only to the conservative contribution. If we use both the conservative and dissipative contributions, we will explicitly make that clear. For all of our numerical results, our $E$ and $J$ values refer to the initial ADM energy and ADM angular momentum of the system. When comparing to analytical results, other works have used the average of the initial and final values, so care should be taken when comparing to previous works.

\section{Numerical Methods}
Simulations were performed using the open-source Einstein Toolkit \cite{einstein_toolkit,NRPyPN:web,AHFinder,Carpet,TwoPunctures} and our initial data and evolution generally follows that of \cite{patrick_paper}, although our specific AMR grid setup and gravitational wave extraction radii vary. We use the open source software Kuibit \cite{kuibit} for post processing our simulations. For the unbound runs, the black holes start on the x-axis at $\pm 30$ $M$. The adaptive mesh refinement (AMR) grid has a half side length of 500 $M$ and consists of eight levels of refinement for a minimum refinement of $\frac{5 \,M}{256}$, with reflection symmetry employed across the x-y plane. No significant difference was found in runs with additional refinement or larger initial separation, and we find that an initial separation of 60 $M$ is sufficient to determine an incoming track before the black holes begin to strongly interact. We replicated a selection of runs from \cite{damour_hopper_scattering} and found our results for the scattering angle were consistent to within our stated uncertainties. 

\begin{figure}
    \centering
    \includegraphics[trim={0 0 0 1cm},clip,width=0.5\textwidth]{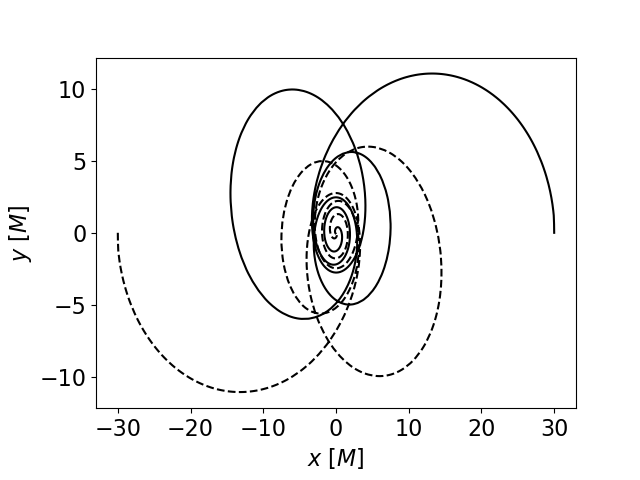}
    \caption{The puncture tracks for a bound orbit of two black holes starting at 60 $M$ separation with $e = 0.75$. Note that the aspect ratio of the figure is roughly 3:1. }
    \label{fig:bound_orbit}
\end{figure}

\begin{figure}
    \centering
    \includegraphics[trim={0 0 0 1cm},clip,width=0.5\textwidth]{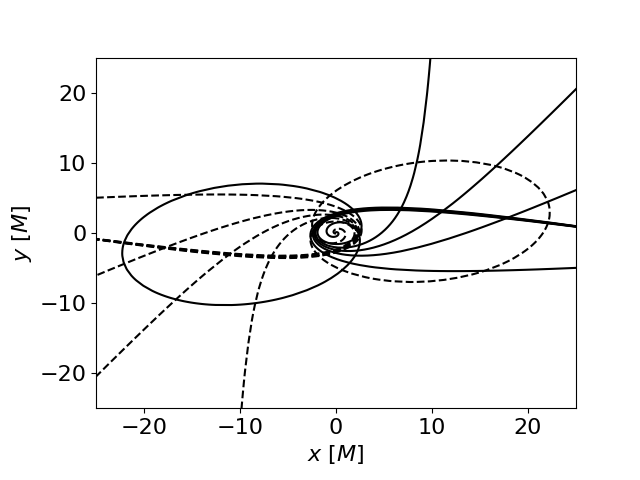}
    \caption{A selection of initially unbound simulations with $E_i$ $\approx$ 1.01525 and varying $J_i$. Included are simulations that, while initially on unbound orbits, become bound due to the emission of radiation. }
    \label{fig:scattering_orbits_E1015}
    \includegraphics[trim={0 0 0 1cm},clip,width=0.5\textwidth]{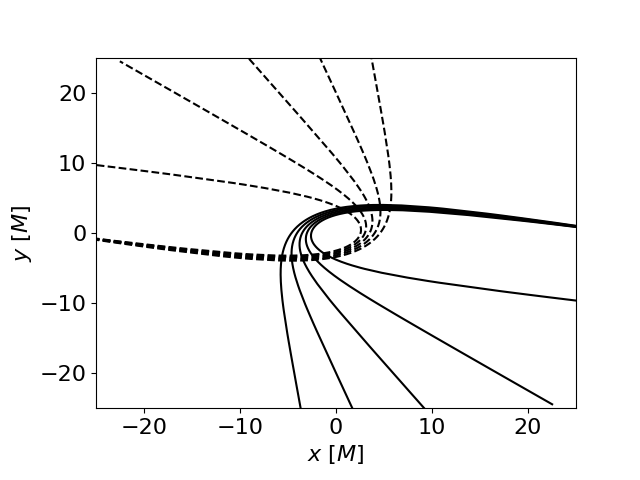}
    \caption{A selection of scattering simulations with $E_i$ $\approx$ 1.0235 and varying $J_i$. All simulations remain unbound after interacting. }
    \label{fig:scattering_orbits_E10235}
\end{figure}

For the bound runs, we quantify our eccentricity as in \cite{RIT_catalog}; we modify the tangential momentum corresponding to a quasicircular momentum $p_{t,qc}$ by a factor $(1 - \alpha)$, i.~e.~$p_t = p_{t,qc} (1-\alpha)$, and then we define our eccentricity as $e = 2\alpha - \alpha^2$. It is important to note that all our bound orbits do not start at the same apocenter, so care must be taken when comparing runs with different initial eccentricities. Figures   ~\ref{fig:bound_orbit}-~\ref{fig:scattering_orbits_E10235} show select examples of bound and unbound orbits.

\subsection{Radiated Quantities}
The accurate extraction of radiated energy and angular momentum from numerical simulations is essential for the work presented here, and significant testing was undertaken to ensure the accuracy of our radiated quantities. Because the waveforms analyzed include those from low eccentricity bound orbits, whose waveforms appear qualitatively similar to traditional quasicircular orbits, and those from both highly eccentric and unbound orbits, whose waveforms consist of a single or a series of isolated pulses, we use both fixed frequency and time domain integration. We extract waveforms at 140 $M$ for initially unbound orbits and at 200 $M$ for bound orbits. We extrapolate our waveforms to infinity using the method presented in \cite{pertubative_extraction}. 
For unbound orbits, there is no physically motivated ``cutoff'' frequency that would be required for fixed frequency integration. Therefore, as was done in \cite{damour_hopper_scattering,Hopper_damour_spinning_scattering}, we resort to time domain integration. 

For the time domain integration, we first remove the junk radiation, which is temporally separated from the physical radiation for our unbound cases. Then, after each integration, we remove a global linear fit for both the real and imaginary components. We do this for all harmonics up to and including the $l = 4$ modes. We find that for even higher-order modes, a significant drift remains in the time domain integration even after removing a global linear fit.  Complications also arise when we consider orbits that are only weakly interacting. These typically correspond to orbits with high initial angular momentum $J \,\gsim\, 1.4\,M^2$. In this weekly interacting regime, there still remain large unphysical drifts in the waveforms after our time domain integration procedure, especially for higher $l$ modes, and we see significant differences between time and frequency domain integrations. Comparison with \cite{damour_hopper_scattering} shows that our simulations are consistent with the lower impact parameter simulations, but we see small but non-negligible deviations for simulations $J \, \gsim\,1.4\,M^2$. Therefore, for our radiative observables we only use runs with $J < 1.4\,{M}^2$, so that the radiated quantities are largely unaffected by the integration choice. We place a conservative error estimate of 5\% on our radiated quantities, based on comparisons of time and frequency domain integrations, comparisons with previously reported values, and comparisons with including both diagonal, $l = |m|$, modes and off diagonal, $l\neq |m|$, modes. Because the physical signal present in off diagonal modes can often be hidden under a significant amount of unphysical noise, we only compare the $l = |m|$ modes when comparing our results to those available through public waveform catalogs.

For all bound waveforms, we use fixed frequency integration with a lower cutoff frequency determined from the corresponding quasicircular orbit \cite{POWER_cite1,POWER_cite2,POWER_cite3}. We compared our results to those obtained from the time integration method described above as well as a full frequency integration, where no cutoff frequency was applied, and found negligible differences for medium and high eccentricity orbits, but did find small differences for low eccentricities. Furthermore, because our low eccentricity runs often require dozens of orbits before merger and our runs starting at large separations take significant amounts of time to complete even one orbit, we do not run all of our simulations until merger since we are only interested in the first orbit. To account for this, we add a windowing function to our waveforms to ensure they smoothly go to zero at both ends. This adds nonphysical features to the end of our waveforms, but this is isolated to the portion of the waveform corresponding to an incomplete orbit, which we do not use. Lastly, we also see the effect of junk radiation being reflected by the outer boundary in our waveforms. This is mainly present in our low eccentricity orbits, and we apply a filter to smooth the data and remove noise for those cases.  We extensively tested different filter and windowing parameters, with both time and frequency domain integration, and found that the differences are, as expected, largest for the low eccentricity waveforms, but even then are negligibly small. We also compared a subset of low eccentricity runs to ones included in the RIT waveform catalog \cite{RIT_catalog} and found differences at the sub percent level for the $l = |m|$ modes, with the differences being largely due to the presence and handling of the initial burst of junk radiation in the respective waveforms. Based on our testing, and to account for the various complicating factors, we place a conservative error estimate of 5\% on our radiated quantities for bound orbits.Tables I-III list the initial parameters and
the derived outputs of each run.
 \begin{table}[!ht]
    \centering
    \renewcommand{\arraystretch}{1.5}
    \begin{tabular}{|c|c|c|c|c|c|c|c|c|c|}
    \hline
        $E$ & $J$ & $p_x$ & $p_y$ & type & $\Delta E$ & $\Delta J$ & $r_{\mathrm{min}} \ [M]$ & $\chi$ [deg] & $\chi_{\rm{error}}$ [deg] \\ \hline
        1.08824 & 1.35 & 0.22387 & 0.02250 & scattering & 7.79017$\times 10^{-2}$ & 4.53266$\times 10^{-1}$ & 2.154 & 17.303 & 0.305 \\ 
        1.06236 & 1.25 & 0.18885 & 0.02083 & scattering & 5.68047$\times 10^{-2}$ & 3.59609$\times 10^{-1}$ & 2.382 & 9.574 & 0.362 \\ 
        1.05569 & 1.225 & 0.17884 & 0.02042 & scattering & 5.00994$\times 10^{-2}$ & 3.28562$\times 10^{-1}$ & 2.487 & 357.418 & 0.383 \\ 
        1.05898 & 1.245 & 0.18382 & 0.02083 & scattering & 4.48590$\times 10^{-2}$ & 3.04176$\times 10^{-1}$ & 2.677 & 303.843 & 0.375 \\ 
        1.03836 & 1.152 & 0.15000 & 0.01920 & unknown & 3.75180$\times 10^{-2}$ & 2.67565$\times 10^{-2}$ & 2.692 & 16.732 & 0.453 \\ 
        1.03237 & 1.125 & 0.13874 & 0.01875 & unknown & 3.37486$\times 10^{-2}$ & 2.48708$\times 10^{-1}$ & 2.762 &   &  \\
        1.04335 & 1.18 & 0.15879 & 0.01967 & scattering & 3.69437$\times 10^{-2}$ & 2.63980$\times 10^{-1}$ & 2.768 & 327.467 & 0.430 \\ 
        1.05248 & 1.225 & 0.17380 & 0.02042 & scattering & 3.95140$\times 10^{-2}$ & 2.76891$\times 10^{-1}$ & 2.790 & 299.096 & 0.396 \\ 
        1.03632 & 1.152 & 0.14625 & 0.01920 & scattering & 3.10330$\times 10^{-2}$ & 2.33308$\times 10^{-1}$ & 2.922 & 322.505 & 0.467 \\ 
        1.07667 & 1.35 & 0.20879 & 0.02250 & scattering & 4.55816$\times 10^{-2}$ & 3.01829$\times 10^{-1}$ & 3.041 & 244.079 & 0.338 \\ 
        1.02504 & 1.1 & 0.12365 & 0.01833 & unknown & 2.44749$\times 10^{-2}$ & 1.98435$\times 10^{-1}$ & 3.083 & 351.854 & 0.554 \\ 
        1.03935 & 1.175 & 0.15174 & 0.01958 & scattering & 2.86661$\times 10^{-2}$ & 2.19993$\times 10^{-1}$ & 3.100 & 286.939 & 0.453 \\ 
        1.03498 & 1.16 & 0.14371 & 0.01933 & scattering & 2.46430$\times 10^{-2}$ & 1.97821$\times 10^{-1}$ & 3.270 & 278.779 & 0.478 \\ 
        1.02207 & 1.1 & 0.11697 & 0.01833 & scattering & 1.79214$\times 10^{-2}$ & 1.59227$\times 10^{-1}$ & 3.497 & 296.827 & 0.595 \\ 
        1.04631 & 1.25 & 0.16368 & 0.02083 & scattering & 2.13688$\times 10^{-2}$ & 1.81150$\times 10^{-1}$ & 3.590 & 213.149 & 0.433 \\ 
        1.01522 & 1.07 & 0.10003 & 0.01783 & scattering & 1.36895$\times 10^{-2}$ & 1.32982$\times 10^{-1}$ & 3.746 & 306.479 & 0.716 \\ 
        1.01519 & 1.07 & 0.09997 & 0.01783 & scattering & 1.36546$\times 10^{-2}$ & 1.32748$\times 10^{-1}$ & 3.749 & 306.208 & 0.718 \\ 
        1.02504 & 1.13 & 0.12357 & 0.01883 & scattering & 1.56213$\times 10^{-2}$ & 1.44677$\times 10^{-1}$ & 3.839 & 256.458 & 0.565 \\ 
        1.01107 & 1.05 & 0.08828 & 0.01750 & unknown & 1.15591$\times 10^{-2}$ & 1.19197$\times 10^{-1}$ & 3.888 &   &  \\
        1.03769 & 1.25 & 0.14855 & 0.02083 & scattering & 1.28360$\times 10^{-2}$ & 1.28834$\times 10^{-1}$ & 3.993 & 183.980 & 0.487 \\ 
        1.01522 & 1.08 & 0.10000 & 0.01800 & scattering & 1.17378$\times 10^{-2}$ & 1.19486$\times 10^{-1}$ & 4.013 & 279.411 & 0.722 \\ 
         1.01107 & 1.055 & 0.08827 & 0.01758 & unknown & 1.06438$\times 10^{-2}$ & 1.12503$\times 10^{-1}$ & 4.025 & 309.859 & 1.547 \\ 
        1.01811 & 1.1 & 0.10745 & 0.01833 & scattering & 1.20071$\times 10^{-2}$ & 1.20529$\times 10^{-1}$ & 4.085 & 261.693 & 0.666 \\ 
        1.01107 & 1.06 & 0.08825 & 0.01767 & scattering & 9.83967$\times 10^{-3}$ & 1.06558$\times 10^{-1}$ & 4.161 & 294.395 & 1.548 \\ 
        1.01524 & 1.1 & 0.10000 & 0.01833 & scattering & 8.97745$\times 10^{-3}$ & 9.99228$\times 10^{-2}$ & 4.366 & 245.710 & 0.732 \\ 
        1.00904 & 1.055 & 0.08188 & 0.01758 & scattering & 8.27826$\times 10^{-3}$ & 9.50840$\times 10^{-2}$ & 4.376 & 292.455 & 1.756 \\ 
        1.01107 & 1.07 & 0.08822 & 0.01783 & scattering & 8.49335$\times 10^{-3}$ & 9.64618$\times 10^{-2}$ & 4.426 & 272.536 & 1.567 \\ 
        1.00790 & 1.05 & 0.07806 & 0.01750 & unknown & 7.72021$\times 10^{-3}$ & 9.09129$\times 10^{-2}$ & 4.445 & 297.853 & 1.907 \\ 
        1.02352 & 1.2 & 0.12000 & 0.02000 & scattering & 6.68706$\times 10^{-3}$ & 8.49546$\times 10^{-2}$ & 4.535 & 181.219 & 0.613 \\ 
        1.01526 & 1.12 & 0.10000 & 0.01867 & scattering & 7.10503$\times 10^{-3}$ & 8.61423$\times 10^{-2}$ & 4.554 & 222.469 & 0.741 \\ 
        1.00790 & 1.055 & 0.07804 & 0.01758 & scattering & 7.16481$\times 10^{-3}$ & 8.64301$\times 10^{-2}$ & 4.579 & 286.691 & 1.915 \\ 
        1.00702 & 1.05 & 0.07500 & 0.01750 & unknown & 6.89075$\times 10^{-3}$ & 8.43555$\times 10^{-2}$ & 4.603 & 295.696 & 2.051 \\ 
        1.05361 & 1.5 & 0.17500 & 0.02500 & scattering & 6.01727$\times 10^{-3}$ & 8.24077$\times 10^{-2}$ & 4.700 & 116.676 & 0.455 \\ 
        1.01529 & 1.14 & 0.10000 & 0.01900 & scattering & 5.75732$\times 10^{-3}$ & 7.55890$\times 10^{-2}$ & 4.741 & 205.302 & 0.750 \\ 
        1.01212 & 1.11 & 0.09122 & 0.01850 & scattering & 5.78263$\times 10^{-3}$ & 7.53606$\times 10^{-2}$ & 4.781 & 224.011 & 0.836 \\ 
        1.00597 & 1.05 & 0.07116 & 0.01750 & unknown & 6.00006$\times 10^{-3}$ & 7.70596$\times 10^{-2}$ & 4.794 & 293.316 & 2.262 \\ 
         \hline
    \end{tabular}
    \label{table:unbound_table1}
    \caption{List of unbound simulations used in this work. Provided are the initial energy, angular momentum, linear momentum, radiated energy, radiated angular momentum, periastron distance, scattering angle, and our error estimate for the scattering angle. Also included are simulations whose end state could not be determined.}
\end{table}
 \begin{table}[!ht]
    \centering
    \renewcommand{\arraystretch}{1.5}
    \begin{tabular}{|c|c|c|c|c|c|c|c|c|c|}
    \hline
        $E$ & $J$ & $p_x$ & $p_y$ & type & $\Delta E$ & $\Delta J$ & $r_{\mathrm{min}}\ [M]$ & $\chi$ [deg] & $\chi_{\mathrm{error}}$ [deg] \\ \hline
       
        1.00716 & 1.06 & 0.07546 & 0.01767 & scattering & 6.07676$\times 10^{-3}$ & 7.75311$\times 10^{-2}$ & 4.828 & 275.236 & 2.042 \\
        1.01530 & 1.152 & 0.10000 & 0.01920 & scattering & 5.11406$\times 10^{-3}$ & 7.03789$\times 10^{-2}$ & 4.854 & 196.649 & 0.756 \\ 
        1.01531 & 1.16 & 0.10000 & 0.01933 & scattering & 4.74490$\times 10^{-3}$ & 6.71961$\times 10^{-2}$ & 4.929 & 191.322 & 0.761 \\ 
        1.00716 & 1.07 & 0.07542 & 0.01783 & scattering & 5.31145$\times 10^{-3}$ & 7.09901$\times 10^{-2}$ & 4.951 & 260.153 & 2.066 \\ 
        1.02352 & 1.26 & 0.11983 & 0.02100 & scattering & 4.17009$\times 10^{-3}$ & 6.14724$\times 10^{-2}$ & 5.019 & 152.558 & 0.634 \\ 
        1.01533 & 1.18 & 0.10000 & 0.01967 & scattering & 4.01742$\times 10^{-3}$ & 6.03967$\times 10^{-2}$ & 5.116 & 179.713 & 0.768 \\ 
        1.01536 & 1.2 & 0.10000 & 0.02000 & scattering & 3.36217$\times 10^{-3}$ & 5.42923$\times 10^{-2}$ & 5.304 & 169.511 & 0.777 \\ 
        1.01536 & 1.2 & 0.10000 & 0.02000 & scattering & 3.41396$\times 10^{-3}$ & 5.44128$\times 10^{-2}$ & 5.304 & 169.910 & 0.776 \\ 
        1.00387 & 1.06 & 0.06276 & 0.01767 & unknown & 3.93181$\times 10^{-3}$ & 5.85889$\times 10^{-2}$ & 5.324 & 277.782 & 2.902 \\ 
        1.00387 & 1.07 & 0.06271 & 0.01783 & scattering & 3.45636$\times 10^{-3}$ & 5.39149$\times 10^{-2}$ & 5.466 & 264.498 & 2.932 \\ 
        1.00457 & 1.08 & 0.06557 & 0.01800 & scattering & 3.35957$\times 10^{-3}$ & 5.29194$\times 10^{-2}$ & 5.480 & 249.814 & 2.712 \\ 
        1.01538 & 1.22 & 0.10000 & 0.02033 & scattering & 2.93103$\times 10^{-3}$ & 4.92161$\times 10^{-2}$ & 5.492 & 161.468 & 0.785 \\ 
        1.02352 & 1.32 & 0.11965 & 0.02200 & scattering & 2.75504$\times 10^{-3}$ & 4.38003$\times 10^{-2}$ & 5.507 & 133.717 & 0.655 \\ 
        1.03819 & 1.5 & 0.14882 & 0.02500 & scattering & 2.72415$\times 10^{-3}$ & 4.41695$\times 10^{-2}$ & 5.616 & 105.391 & 0.546 \\ 
        1.03501 & 1.5 & 0.14287 & 0.02500 & scattering & 2.27897$\times 10^{-3}$ & 4.03050$\times 10^{-2}$ & 5.858 & 103.717 & 0.570 \\ 
        1.02352 & 1.38 & 0.11946 & 0.02300 & scattering & 1.96632$\times 10^{-3}$ & 3.81475$\times 10^{-2}$ & 6.000 & 119.804 & 0.669 \\ 
        1.00287 & 1.1 & 0.05818 & 0.01833 & scattering & 2.11608$\times 10^{-3}$ & 3.95836$\times 10^{-2}$ & 6.092 & 241.420 & 3.488 \\ 
        1.03195 & 1.5 & 0.13692 & 0.02500 & scattering & 1.89240$\times 10^{-3}$ & 3.51536$\times 10^{-2}$ & 6.117 & 102.382 & 0.597 \\ 
        1.02902 & 1.5 & 0.13096 & 0.02500 & scattering & 1.56783$\times 10^{-3}$ & 3.05968$\times 10^{-2}$ & 6.392 & 101.379 & 0.627 \\ 
        1.02352 & 1.44 & 0.11926 & 0.02400 & scattering & 1.42527$\times 10^{-3}$ & 2.88249$\times 10^{-2}$ & 6.497 & 109.057 & 0.684 \\ 
        1.00719 & 1.2 & 0.07500 & 0.02000 & scattering & 1.44896$\times 10^{-3}$ & 3.16453$\times 10^{-2}$ & 6.521 & 174.055 & 2.306 \\ 
        1.02621 & 1.5 & 0.12501 & 0.02500 & scattering & 1.27425$\times 10^{-3}$ & 2.47545$\times 10^{-2}$ & 6.685 & 100.764 & 0.658 \\ 
        1.02620 & 1.5 & 0.12500 & 0.02500 & scattering & 1.23879$\times 10^{-3}$ & 2.25880$\times 10^{-2}$ & 6.686 & 100.806 & 0.658 \\ 
        1.00597 & 1.2 & 0.07050 & 0.02000 & scattering & 1.24773$\times 10^{-3}$ & 2.87740$\times 10^{-2}$ & 6.777 & 177.073 & 2.584 \\ 
        1.00127 & 1.152 & 0.05000 & 0.01920 & scattering & 9.79912$\times 10^{-4}$ & 2.46117$\times 10^{-2}$ & 7.316 & 226.853 & 4.950 \\ 
        1.03934 & 1.8 & 0.15000 & 0.03000 & scattering & 9.18176$\times 10^{-4}$ & 1.81959$\times 10^{-2}$ & 7.545 & 73.300 & 0.576 \\ 
        1.01623 & 1.5 & 0.10120 & 0.02500 & scattering & 5.75278$\times 10^{-4}$ & 1.39734$\times 10^{-2}$ & 8.086 & 102.838 & 1.681 \\ 
        1.00133 & 1.2 & 0.05000 & 0.02000 & scattering & 6.42959$\times 10^{-4}$ & 1.90109$\times 10^{-2}$ & 8.088 & 207.658 & 5.130 \\ 
        1.01212 & 1.5 & 0.08965 & 0.02500 & scattering & 3.94042$\times 10^{-4}$ & 1.16332$\times 10^{-2}$ & 8.932 & 105.838 & 2.060 \\ 
        1.02352 & 1.74 & 0.11815 & 0.02900 & scattering & 3.64771$\times 10^{-4}$ & 7.92303$\times 10^{-3}$ & 9.046 & 78.241 & 1.490 \\ 
        1.02352 & 1.8 & 0.11790 & 0.03000 & scattering & 3.24678$\times 10^{-4}$ & 8.44055$\times 10^{-3}$ & 9.566 & 73.381 & 1.543 \\ 
        1.00762 & 1.5 & 0.07500 & 0.02500 & scattering & 2.34247$\times 10^{-4}$ & 9.00104$\times 10^{-3}$ & 10.220 & 113.201 & 2.812 \\ 
        1.00597 & 1.5 & 0.06889 & 0.02500 & scattering & 1.86569$\times 10^{-4}$ & 7.78071$\times 10^{-3}$ & 10.846 & 118.307 & 3.285 \\ 
        1.01536 & 1.8 & 0.09747 & 0.03000 & scattering & 1.52700$\times 10^{-4}$ & 4.69860$\times 10^{-3}$ & 11.344 & 77.747 & 2.150 \\ 
        1.02353 & 2.04 & 0.11681 & 0.03400 & scattering & 1.55428$\times 10^{-4}$ & 4.51851$\times 10^{-3}$ & 11.674 & 61.077 & 1.819 \\ 
        1.00176 & 1.5 & 0.05000 & 0.02500 & scattering & 8.86609$\times 10^{-5}$ & 4.89418$\times 10^{-3}$ & 13.206 & 146.075 & 6.323 \\ 
        1.00597 & 1.8 & 0.06686 & 0.03000 & scattering & 4.59814$\times 10^{-5}$ & 2.50142$\times 10^{-3}$ & 15.288 & 92.987 & 4.440 \\ \hline
\end{tabular}
\label{table:unbound_table2}
\caption{Continued list of unbound simulations}
\end{table}
\begin{table}[!ht]
\label{table:bound_table1}
    \centering
    \renewcommand{\arraystretch}{1.5}
    \begin{tabular}{|c|c|c|c|c|c|c|c|c|}
    \hline
        $E$ & $J$ & ecc & initial separation [$M$] & $\Delta E$ & $\Delta J$ & $r_{\mathrm{min}}\ [M]$ & $\mathrm{T} \ [M]$& $\Delta \Phi$ [deg]\\ \hline
        0.98986 & 0.91178 & 0.319 & 15 & 4.42046$\times 10^{-3}$ & -7.30813$\times 10^{-2}$ & 5.019 & 261.372 & 236.477 \\  
        0.99099 & 0.99467 & 0.190 & 15 & 6.73150$\times 10^{-4}$ & -2.26849$\times 10^{-2}$ & 8.500 & 354.730 & 125.441 \\  
        0.99180 & 1.04993 & 0.098 & 15 & 2.94668$\times 10^{-4}$ & -1.40675$\times 10^{-2}$ & 11.269 & 408.938 & 93.793 \\  
        0.99222 & 1.07756 & 0.049 & 15 & 2.05630$\times 10^{-4}$ & -1.13047$\times 10^{-2}$ & 12.896 & 436.078 & 78.158 \\  
        0.99310 & 1.01531 & 0.437 & 25 & 9.51433$\times 10^{-4}$ & -2.55916$\times 10^{-2}$ & 7.281 & 513.844 & 131.100 \\  
        0.99329 & 1.04916 & 0.399 & 25 & 5.75235$\times 10^{-4}$ & -1.90206$\times 10^{-2}$ & 8.413 & 549.141 & 111.921 \\  
        0.99350 & 1.08300 & 0.360 & 25 & 3.63865$\times 10^{-4}$ & -1.46049$\times 10^{-2}$ & 9.615 & 557.578 & 91.216 \\  
        0.99371 & 1.11684 & 0.319 & 25 & 2.38527$\times 10^{-4}$ & -1.15290$\times 10^{-2}$ & 10.909 & 590.484 & 80.117 \\  
        0.99392 & 1.15069 & 0.277 & 25 & 1.60519$\times 10^{-4}$ & -9.22901$\times 10^{-3}$ & 12.320 & 623.531 & 70.566 \\  
        0.99405 & 1.17099 & 0.252 & 25 & 1.28010$\times 10^{-4}$ & -8.14328$\times 10^{-3}$ & 13.231 & 644.063 & 65.438 \\  
        0.99437 & 1.21837 & 0.190 & 25 & 7.78028$\times 10^{-5}$ & -6.18332$\times 10^{-3}$ & 15.578 & 698.484 & 55.655 \\  
        0.99461 & 1.25222 & 0.144 & 25 & 5.56534$\times 10^{-5}$ & -5.07545$\times 10^{-3}$ & 17.478 & 741.797 & 49.296 \\  
        0.99485 & 1.28606 & 0.097 & 25 & 4.07362$\times 10^{-5}$ & -4.30982$\times 10^{-3}$ & 19.604 & 793.828 & 44.707 \\  
        0.99582 & 1.01191 & 0.698 & 50 & 2.02423$\times 10^{-3}$ & -3.93266$\times 10^{-2}$ & 5.993 & 886.563 & 147.130 \\  
        0.99604 & 1.14989 & 0.609 & 50 & 3.43879$\times 10^{-4}$ & -1.32908$\times 10^{-2}$ & 9.690 & 1141.453 & 81.924 \\  
        0.99629 & 1.28788 & 0.510 & 50 & 8.93953$\times 10^{-5}$ & -5.92305$\times 10^{-3}$ & 13.999 & 1299.656 & 55.464 \\  
        0.99640 & 1.00104 & 0.750 & 60 & 2.75363$\times 10^{-3}$ & -4.77044$\times 10^{-2}$ & 5.526 & 1007.803 & 160.989 \\  
        0.99771 & 1.02019 & 0.840 & 100 & 2.60412$\times 10^{-3}$ & -4.56601$\times 10^{-2}$ & 5.714 & 1785.656 & 150.046 \\  
        0.99774 & 1.08396 & 0.819 & 100 & 1.09557$\times 10^{-3}$ & -2.65850$\times 10^{-2}$ & 7.315 & 2123.016 & 109.200 \\  
        0.99777 & 1.14772 & 0.798 & 100 & 5.41450$\times 10^{-4}$ & -1.73048$\times 10^{-2}$ & 8.798 & 2385.281 & 86.021 \\ \hline
    \end{tabular}
    \caption{List of bound simulations used in this work. Provided are the initial energy, angular momentum, approximate eccentricity, initial separation as well as the periastron distance, radiated energy, radiated angular momentum, period and periastron advance of the first orbit.}
\end{table}
\subsection{Scattering Angle}
To compute the scattering angle of each simulation, we follow a procedure similar to \cite{damour_hopper_scattering}. First, we convert the Cartesian coordinates of the black hole, provided through the AHFinderDirect thorn \cite{AHFinder}, to polar coordinates ($r,\phi$). We then define an incoming and outgoing track for which we define an incoming and outgoing angle. For the incoming track we choose the portion of the simulation in which the black holes travel from $r = 55$ $M$ to 25 $M$. For the outgoing track we start at 25 $M$ and choose the largest final separation such that $100\,M \,\geq r_f\, \geq\, 60\,M$. These values were chosen such that all simulations reach a separation of at least 60 $M$, but we are able to use additional data from certain simulations while staying in the same refinement zone. 
For each track we fit the angle to a polynomial of degree $n$ in $1/r$, where $n\,\leq\, 5$, and extrapolate to infinity. We then choose the lowest $n$ such that the sum of the squares of the residuals is less than $10^{-5}$. We calculate an incoming and outgoing angle and define the final scattering angle as 
\beq
    \chi = \chi_{\mathrm{in}} - \chi_{\mathrm{out}} - \pi.
\eeq
Since we can also define an incoming angle using the momenta defined in the initial conditions, we define the error in our scattering angle as the difference between the angle determined for the incoming track through the fitting procedure and the initial momentum. For the vast majority of simulations, these two methods agree very well. However, for the weakest interacting simulations, we start to see deviations between these two methods, with the largest deviation being around 6 degrees. For the scattering angle we do not place any restrictions on the maximum initial angular momentum like we do for our radiated quantities.
\subsection{Dynamical Invariants}
The two dynamical invariants we obtain from numerical simulations are the apocenter period and the periastron advance. The apocenter period is relatively straightforward to obtain; we simply identify the times of maximum radii in our orbits and find the difference in the times. We note that while the value of the maximum radii is gauge dependent, the times at which they occur is relatively insensitive to gauge choices. For the periastron advance, we convert our separation vector into polar coordinates, and compute the change in the angle at our apocenter times. 
\section{The end state of initially unbound orbits}
\begin{figure}
    \centering
    \includegraphics[trim={0 0 0 1cm},clip,width=0.5\textwidth]{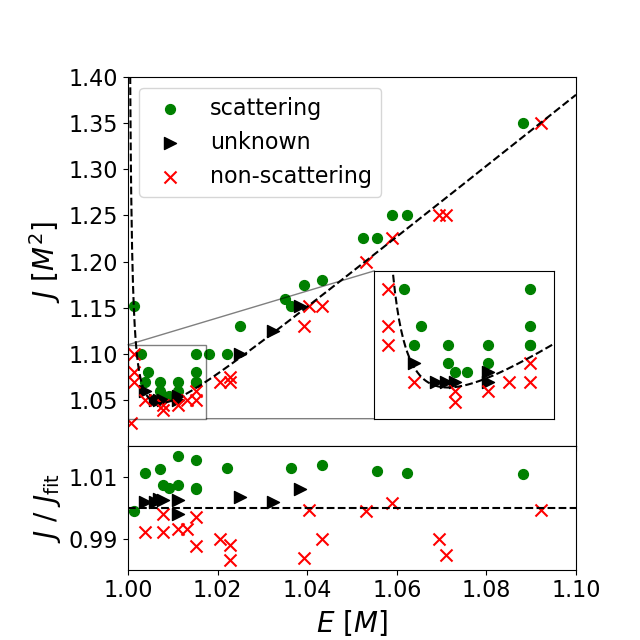}
    \caption{Top: A selection of initially unbound runs with varying end states. The dashed line represents the approximate separatrix between scattering and non-scattering orbits obtained by interpolating to the $\varepsilon = 0$ surface and fitting the result to the form given in Eq.~\eqref{eq:separatrix_fit}.\newline
    Bottom: Ratio of initial angular momentum of the run to the angular momentum predicted by the fit for select runs, with the dashed line indicating an equal ratio.}
    \label{fig:scattering_limits}
\end{figure}

 Since the orbital energy and angular momentum of a system is not a conserved quantity, but decreases over time due to the emission of gravitational waves, a system that is initially on an unbound trajectory may emit enough radiation to become bound. Previous works \cite{strong_field_scattering,strong_field_spinning_scattering,Pretorius_boundary,boundary_sperhake,zoom_whirl_sperhake} have studied the boundary between orbits that remain unbound and those that become bound. The authors of \cite{Pretorius_boundary} noted the presence of a unique class of orbits near the boundary that exhibit zoom-whirl behavior. These type of orbits are a critical phenomenon that are exponentially sensitive to the initial conditions. Therefore, obtaining zoom-whirl behavior amounts to fine tuning initial conditions near the boundary of scattering and non-scattering orbits. While previous works have studied the boundary at specific angular momentum or energies, to the author's knowledge no work has produced an approximate boundary across a wide range of initial energies and angular momenta. Fig.~\ref{fig:scattering_limits} shows a selection of orbits, all of which start on unbound orbits, that either remain unbound or become bound due to the emission of radiation. We determine whether an orbit is unbound or bound based on whether it's binding energy $\varepsilon$ remains greater than zero after the two black holes separate. It is important to note that a visual check is not sufficient as there remain orbits that appear to remain unbound, but are in fact bound with very long orbital periods. Therefore, the radiation provides a more rigorous approach to classifying the end state of initially unbound orbits. Using our unbound and unknown data points, we then interpolate to the $\varepsilon = 0$ surface and fit a function of the form of Eq.~\eqref{eq:separatrix_fit} to the result. Through this procedure, we obtain an approximate separatrix of the form
 \bea
  \label{eq:separatrix_fit}
 J(E)_{\varepsilon=0} &\approx& \frac{a(E/M)^2 + b E/M + c}{E/M-1}\,M^2\,,\\
 a &=& 3.8733\,,  \nonumber \\
 b &=& -6.7554\,, \nonumber \\
 c &=& 2.8823\,.  \nonumber 
 \eea
 As shown in Fig.~\ref{fig:scattering_limits}, this equation splits through the unbound and bound data points at lower initial $E$ and $J$ values, but slightly deviates into the non-scattering space at higher energies and as the initial energy approaches the parabolic limit. Nonetheless, for future work studying the boundary between scattering and non-scattering orbits, Eq.~\eqref{eq:separatrix_fit} should decrease the initial guesswork needed to approach the boundary. Eq.~\eqref{eq:separatrix_fit} can also be used in resummation procedures for the scattering angle such as the one proposed in \cite{B2B_paper1,strong_field_scattering}, and we do so in Sec.~\ref{sec:dyn}.
 
 Studying the behavior of the separatrix as a function of energy, we can see that at any given angular momentum, there are two boundary points discriminating scattering from non-scattering orbits. This is due to the fact that, at a given angular momentum, the radiated energy increases non linearly as a function of the initial energy. Let us consider the example of different orbits all at the same initial angular momentum $J_0$. The first type of orbit has an initial binding energy only slightly above that needed to be on an initially unbound orbit. The system radiates enough energy that the initially unbound system becomes bound. As we increase the initial energy of the orbit, this increase outpaces the resulting increase in radiated energy. The second type of orbit has enough initial energy that it is able to remain unbound. As we continue to increase the initial energy of the orbits beyond this point, the increase in radiated energy begins to outpace the increase in initial energy. The third type of orbit, with still higher initial energy, now radiates enough energy that the system becomes bound once again. Increasing the energy beyond this point simply decreases the time between the initial interaction and the eventual merger. We also note that as we go to higher initial $E$ and $J$ values, the parameter space of initial energies and angular momentum values allowing for large bound orbits becomes smaller and the transition from scattering orbits to a delayed merger to a prompt merger becomes much smaller.
\section{The Boundary to Bound State Relations}
Here, we present the results of testing the radiative and orbital relationships presented in the B2B papers \cite{B2B_paper1,B2B_paper2,B2B_paper3}. We focus on the energy and angular momentum radiated during one bound orbit as well as the period and periastron advance. For simplicity, we only use the first orbit from our bound runs.

The B2B relations in the conservative sector rely on the knowledge that the distance of closest approach for an unbound orbit, $\tilde{r}_{-}$, can be linked to the apocenter, $r_{+}$, and pericenter, $r_{-}$, of a bound orbit through analytic continuation. That is, 
\bea
\label{eq:endpoints}
r_{-} (J,\varepsilon) &=& \tilde{r}_{-} (J,\varepsilon)\\
r_{+} (J,\varepsilon) &=& \tilde{r}_{-} (-J,\varepsilon)
\eea
where $\varepsilon < 0$ and $J > 0\,M^2$.

As was shown in \cite{B2B_paper3}, assuming the adiabatic approximation, we can relate the energy and angular momentum radiated in unbound orbits to those of bound orbits via analytic continuation. An inherent limitation in connecting bound and unbound orbits is that unbound orbits, unlike bound orbits, have a minimum angular momentum constraint, as can be seen in Fig.~\ref{fig:scattering_limits}. We find that when $J \lsim \,1.05 \,M^2$, we are unable to obtain unbound orbits and it is unclear how well we can model bound orbits beyond this limit. In this work we choose to mostly limit our bound orbits to $J\, \geq\, M^2$, with two exceptions where we go slightly below this limit for orbits starting at 15 $M$ separation. Due to this limitation, we are unable to test the full range of eccentricities starting at one separation. Therefore, we adopt various starting separations for our bound orbits, allowing us to test a wide array of eccentricities without having to significantly lower our angular momentum below the scattering limit. 

The main purpose of this work is to determine whether the B2B relationships hold in full GR, where both conservative and dissipative effects are at play. In order to test the results, we need to generate a general functional form for the observables as a function of energy and angular momentum. Our functional forms consist of two parts. The first part contains analytical information that has been shown to follow the B2B relationships. The second part is a fitting term, calibrated to NR simulations.
\subsection{Radiative Observables}
We begin by testing the equation for the energy and angular momentum radiated in one bound orbit introduced in \cite{B2B_paper3},
\bea
\label{eq:B2B_paper3}
    \centering
        \Delta E_{\mathrm{ell}} (E,J) &=& \Delta E_{\mathrm{hyp}}(E,J) - \Delta E_{\mathrm{hyp}}(E,-J) \\
        \Delta J_{\mathrm{ell}} (E,J) &=& \Delta J_{\mathrm{hyp}}(E,J) + \Delta J_{\mathrm{hyp}}(E,-J)
\eea
where $E < M$ and $J > 0\,M^2$.

For the analytical information going into our functional form, we choose to use the 3PN conservative equations for radiated energy and angular momentum, derived in \cite{3PN_hyperbolic,3PN_elliptical} and shown to follow the B2B relations in \cite{B2B_paper3}. These expressions include even and odd terms up to $1/j^{13}$. We will refer to the 3PN conservative equations simply as 3PN.

While the B2B relations can be used in either direction, assuming we do not include both even and odd terms in our fitting function, here we choose to use bound data to predict unbound data. We do so primarily for clarity of presentation. We have tested both directions, and our overall findings remain the same. On the bound side we place the constraint that the radiated quantity coming from the fit should not deviate from the simulation result by more than 10\% for any data point. A significant problem arises from attempting to fit all of our scattering data to a single fitting term. Additionally, while we could choose to make our fitting form include both odd and even terms, for example an even function of angular momentum for the radiated energy, doing so would make it difficult to assess the accuracy of the B2B relations as it would leave a free parameter that is not constrained by applying the B2B relations.

Because the B2B relations may only apply within a certain region of the parameter space, we test them on subsets of our data; we start with our scattering orbits that have the smallest periastron distance, and successively add in orbits with larger periastron distances in increments of $0.25 \,M$. We then repeat this procedure in the opposite direction. The periastron distances are gauge dependent, but since we are only interested in roughly categorizing the ``strongness" of the interactions, we can nonetheless use it to group the unbound runs.

Noting the sign flip between the equations for radiated energy and angular momentum, we choose an even function for our angular momentum equation and an odd function for our energy equation. For bound orbits with $e\,\gsim \, 0.5$, we find that the functional form
\beq
    \label{eq:energy_fitting_form}
    \Delta X_{\mathrm{ell}_{\mathrm{NR}}} = \frac{a  (e^2_\mathrm{N})^b}{j^{c}}
\eeq
where $X\equiv\{E,J\}$, provides excellent fits to the numerical data on the bound side. Our choice of $c = 15$ for the radiated energy and $c = 14$ for the radiated angular momentum was chosen due to the analytical 3PN result including terms up to $j^{13}$. Similar or worse results were found for other reasonable choices of the exponent. Because $e^2_\mathrm{N}$ can take negative values for bound orbits, we only allow integer values for $b$. We set the value of $b$ by directly fitting the bound data. Fitting the radiated quantity in a bound orbit to the functional form
\beq
\label{eq:energy_fitting_PN_and_NR}
    \Delta X_{\mathrm{ell}} = \Delta X_{\mathrm{ell}_{\mathrm{3PN}}} + \frac{a(e^2_\mathrm{N})^b}{j^{c}}
\eeq
where $X\equiv\{E,J\}$, $ c \equiv\{15,14\}$ and $b = 1$ can effectively describe bound orbits with eccentricity $e\,\gsim \, 0.5$. In our data set, this corresponds to starting apocenters of 50, 60 and 100 M. Incorporating our fitting term, the B2B relation for radiated energy reads
\bea
\label{eq:B2B_relation_full}
    \Delta E_{\mathrm{ell}_{\mathrm{3PN}}} (\varepsilon,j) &+& \frac{a_{\mathrm{ell}} e^2_\mathrm{N}}{j^{15}} = \nonumber \\
    \Delta E_{\mathrm{hyp}_{\mathrm{3PN}}}(\varepsilon,j) &-& \Delta E_{\mathrm{hyp}_{\mathrm{3PN}}}(\varepsilon,-j) + \frac{2 a_{\mathrm{hyp}} e^2_\mathrm{N}}{j^{15}}
\eea
where $J > 0 $ and $\varepsilon < 0$.
Similarly, the B2B relation for radiated angular momentum reads
\bea
\label{eq:B2B_relation_full_angmom}
    \Delta J_{\mathrm{ell}_{\mathrm{3PN}}} (\varepsilon,j) &+& \frac{a_{\mathrm{ell}} e^2_\mathrm{N}}{j^{14}} = \nonumber \\
    \Delta J_{\mathrm{hyp}_{\mathrm{3PN}}}(\varepsilon,j) &+& \Delta J_{\mathrm{hyp}_{\mathrm{3PN}}}(\varepsilon,-j) + \frac{2 a_{\mathrm{hyp}} e^2_\mathrm{N}}{j^{14}}
\eea
where $J > 0 $ and $\varepsilon < 0$.
For a fitting term of the form of Eqs.~\eqref{eq:B2B_relation_full} and~\eqref{eq:B2B_relation_full_angmom}, the B2B relations require
\beq
\label{eq:B2B_ell_hyp_relation}
    a_{\mathrm{hyp}} = \frac{a_{\mathrm{ell}}}{2}.
\eeq

 We already know that the analytical hyperbolic information follows Eq.~\eqref{eq:B2B_paper3}, so that
\bea
    \Delta E_{\mathrm{hyp}_{\mathrm{3PN}}} (J) - \Delta E_{\mathrm{hyp}_{\mathrm{3PN}}} (-J) = \Delta E_{\mathrm{ell}_{\mathrm{3PN}}} (J),\\
    \Delta J_{\mathrm{hyp}_{\mathrm{3PN}}} (J) + \Delta J_{\mathrm{hyp}_{\mathrm{3PN}}} (-J) = \Delta J_{\mathrm{ell}_{\mathrm{3PN}}} (J).
\eea
Thus, by obtaining a value for $a_{ell}$ by fitting directly to the bound data, we can obtain the B2B predicted value for $a_{hyp}$ for the respective radiated quantity. This then gives us the B2B predicted radiated quantity for the unbound side, which we can directly compare to our unbound NR results.

\begin{figure}
    \centering
    \includegraphics[trim={0 0 0 1cm},clip,width=0.5\textwidth]{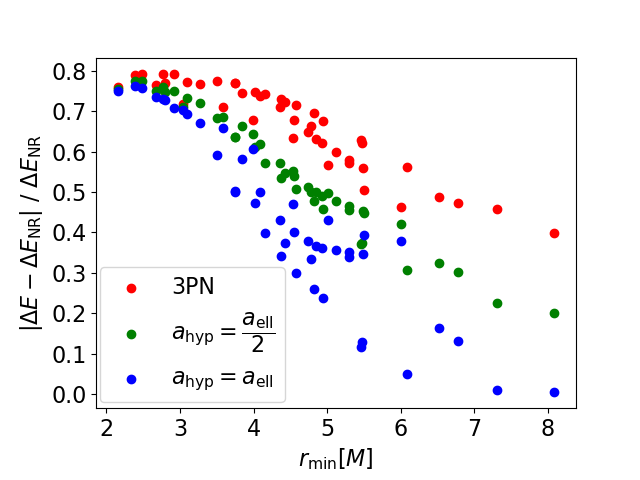}
    \caption{Mean fractional difference from NR for the radiated energy of unbound encounters obtained from applying the B2B relations on bound data with $e > 0.5$. We compare using the B2B relations on a 3PN estimate for radiated energy from a bound orbit and an NR-calibrated fit to the bound data consisting of 3PN and a fitting term. The B2B relation implies $a_{\mathrm{hyp}} = \frac{a_{\mathrm{ell}}}{2}$ in Eq. \eqref{eq:B2B_relation_full}, but we also test $a_{\mathrm{hyp}} = a_{\mathrm{ell}}$.}
    \label{fig:compare_B2B_arb_E}
    \includegraphics[trim={0 0 0 1cm},clip,width=0.5\textwidth]{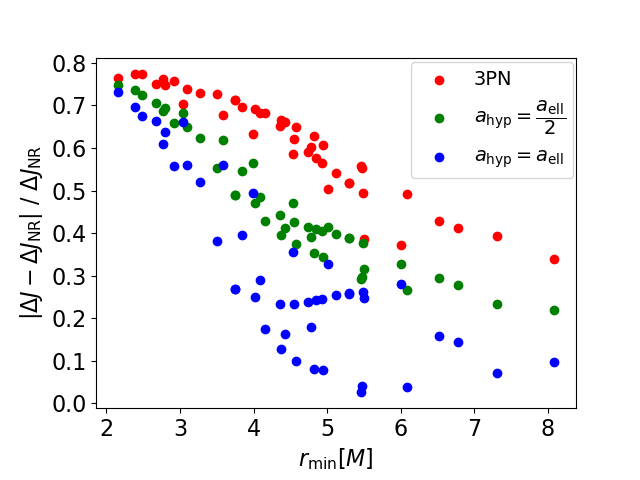}
    \caption{Same as Fig.~\ref{fig:compare_B2B_arb_E}, but comparing radiated angular momentum, $\Delta J$.}
    \label{fig:compare_B2B_arb_J}
\end{figure}

We emphasize again that all references to 3PN in this work refer only to the conservative contributions, and not dissipative contributions. As a very simple first test, we can take the best fitting elliptical coefficient $a_{\mathrm{ell}}$, and apply the B2B relationship to determine the corresponding scattering radiative observable. For comparison purposes, we also include a fit where we arbitrarily change the B2B relationship such that $a_{\mathrm{hyp}} = a_{\mathrm{ell}}$. As can be seen in Figs.~\ref{fig:compare_B2B_arb_E} and \ref{fig:compare_B2B_arb_J}, while the B2B predicted value shows a clear improvement over the 3PN analytic relationship, arbitrarily doubling the fitting coefficient yields even better results, suggesting that the improvement may be spurious.

\begin{figure}
    \centering
    \includegraphics[trim={0 0 0 1cm},clip,width=0.5\textwidth]{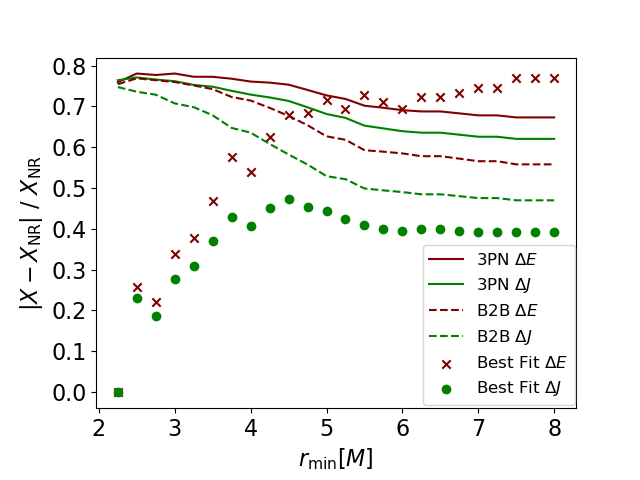}
    \caption{Mean fractional difference from NR for the radiated energy and angular momentum of unbound encounters, $X=\{\Delta E,\Delta J\}$ using either 3PN bound data for runs with $e > 0.5$, 3PN and a fitting term (``Best Fit''), or the B2B predicted unbound result. $r_{\mathrm{min}}$ indicates the maximum periastron distance from all of the unbound runs used for the respective data point. We start with only the strongest interacting runs and add additional runs in increments of 0.25 $M$.}
    \label{fig:max_rad_compare_ecc_gtr_0.5}
    \includegraphics[trim={0 0 0 1cm},clip,width=0.5\textwidth]{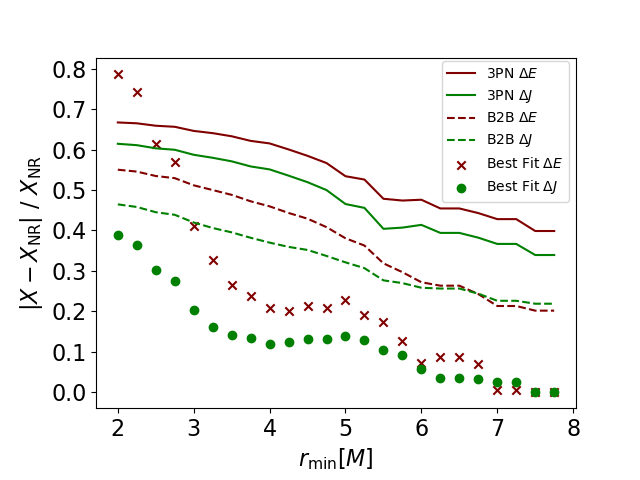}
    \caption{Same as Fig.~\ref{fig:max_rad_compare_ecc_gtr_0.5}, except $r_{\mathrm{min}}$ now indicates the minimum periastron distance from all of the unbound runs use for the respective data point. We start with only the weakest interacting runs and add additional runs in increments of 0.25 $M$.}
    \label{fig:min_rad_compare_ecc_gtr_0.5}
\end{figure}

\begin{figure}
    \centering
    \includegraphics[trim={0 0 0 1cm},clip,width=0.5\textwidth]{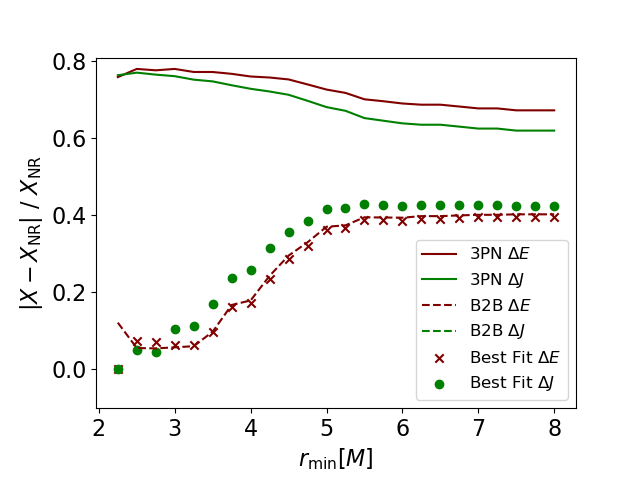}
    \caption{Same as Fig. \ref{fig:max_rad_compare_ecc_gtr_0.5} but only using data from bound orbits starting at 25 $M$ separation The B2B predicted value for the radiated angular momentum and is not visible in this plot due to its large value.}
    \label{fig:max_rad_compare_25M}
    \includegraphics[trim={0 0 0 1cm},clip,width=0.5\textwidth]{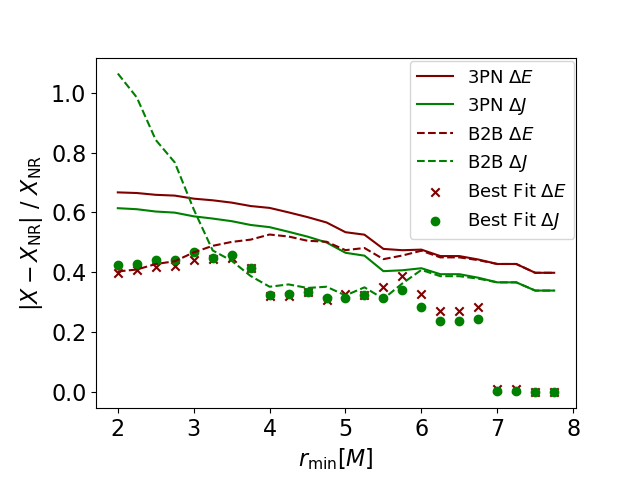}
    \caption{Same as Fig.~\ref{fig:min_rad_compare_ecc_gtr_0.5} but only using data from bound orbits starting at 25 $M$ separation.}
    \label{fig:min_rad_compare_25M}
\end{figure}

\begin{figure}
    \centering
    \includegraphics[trim={0 0 0 1cm},clip,width=0.5\textwidth]{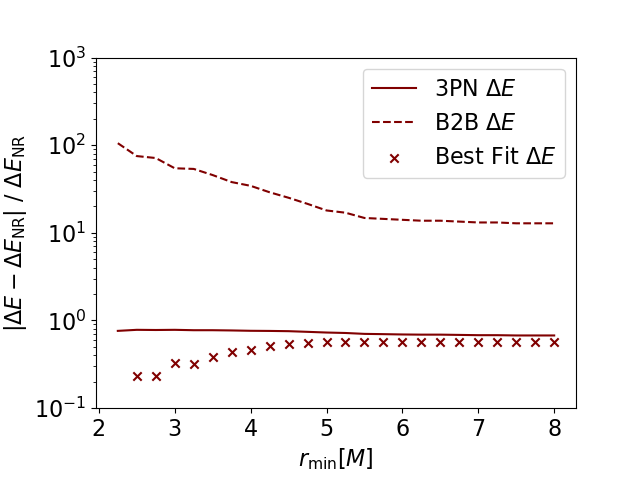}
    \caption{Same as Fig.~\ref{fig:max_rad_compare_25M} but using an alternate fitting form where we set $b = 6$ in Eq.~\eqref{eq:eps_form}.}
    \label{fig:E_rad_compare_25M_eps6}
\end{figure}

It is important to note that in Figs.~\ref{fig:compare_B2B_arb_E}--\ref{fig:E_rad_compare_25M_eps6}, each data point corresponds to the maximum or minimum periastron distance of the multiple unbound runs used for that data point. For each data point, we use all of our unbound runs that have a periastron distance smaller or larger, depending on the figure, than the indicated $r_\text{min}$ value on the x-axis.

To further test the B2B equations, we allow $a_{\mathrm{hyp}}$ to be a free parameter and fit directly to the NR unbound results using a form similar to Eq. \eqref{eq:energy_fitting_PN_and_NR}, but using the unbound 3PN equation for the analytical term.\
\beq
\label{eq:energy_fitting_hyp_PN_and_NR}
    \Delta X_{\mathrm{hyp}} = \Delta X_{\mathrm{hyp}_{\mathrm{3PN}}} + \frac{a(e^2_\mathrm{N})^b}{j^{c}}
\eeq
where $X\equiv\{E,J\}$, $ c \equiv\{15,14\}$, $b = 1$ and $a$ is a free parameter. This allows us to determine what is the theoretical best result we can expect from our chosen fitting form.

From Figs.~\ref{fig:max_rad_compare_ecc_gtr_0.5} and \ref{fig:min_rad_compare_ecc_gtr_0.5}, we can see that when we fit our functional form for the unbound side, given in Eq.~\eqref{eq:energy_fitting_hyp_PN_and_NR}, directly to our unbound NR data we observe significant improvements when we isolate either the weak or strong field data. When instead, we first fit the bound data and follow the B2B relations given by Eqs. \eqref{eq:B2B_relation_full} - \eqref{eq:B2B_ell_hyp_relation} to obtain the unbound result, we do obtain improvements over using only the 3PN data, but there still remains significant disagreement with the NR data, and the B2B result performs significantly worse than the free parameter result. 

To test low eccentricity bound orbits, we split our data into those starting at an apocenter of 15 $M$ and 25 $M$. In order to allow the free parameter to be in the range of the B2B predicted value, we change our functional form to
\beq
    \Delta X_{\mathrm{ell}} = \Delta X_{\mathrm{ell_{3PN}}} + a_{\mathrm{ell}}\frac{\varepsilon^b}{j^c}\,.
    \label{eq:eps_form}
\eeq
For bound orbits starting at 25 $M$, we set $b = 4$, $c = 17$ for the radiated energy and $b = 4$, $c = 14$ for the radiated angular momentum. We follow the same strategy as was done for the previous functional form for transferring bound and unbound data. Fig.~\ref{fig:max_rad_compare_25M} shows that when we isolate the strong field scattering, the B2B relation for the radiated energy not only matches closely to the best fit value, but also provides significant improvements over the analytical 3PN estimate. While this could suggest that higher order PN terms can be well captured by our NR fitting term, and that perhaps the B2B relation holds in the dissipative sector as well, we believe this result does not have any physical significance. This is mainly due to the fact that we can produce many different NR fitting terms, that give very good results when $a_{\mathrm{hyp}}$ is allowed to be a free parameter, but terrible results when we restrict $a_{\mathrm{hyp}}$ according to the B2B relations. For example, if we set $b = 6$ for the radiated energy, Fig. ~\ref{fig:E_rad_compare_25M_eps6} shows that the free parameter result still yields an improvement over the PN result, but the B2B result is completely incorrect. 

We therefore believe any agreement between B2B and our simulation results are a coincidence resulting from our specific choice of the fitting term. This stresses an important point, that while it is likely there exists some NR fitting term that provides results consistent with the B2B relations for any selection of bound runs, the large number of alternate fitting terms that do not work with the B2B relations indicates that this is more likely to be coincidence than of any physical significance. Replicating this procedure for bound runs starting at 15 $M$ separation also showed no evidence of the B2B relations working in the dissipative regime.
\subsection{Dynamical Invariants}
\label{sec:dyn}
\subsubsection{Scattering Angle}
We now compare the scattering angle obtained from our numerical simulations to existing analytical information in the PN and PM expansions. We compare our numerically obtained scattering angles to the PN expanded analytical estimates provided in \cite{B2B_paper3}, which was obtained through derivations provided in \cite{PN_scattering_paper1,PN_scattering_paper2,PN_scattering_paper3,PN_scattering_paper4,PN_scattering_paper5,PN_scattering_paper6,PN_scattering_paper7}, and contains terms up to 4PN that were shown to follow the B2B relations. Since the B2B relations face limitations when including nonlocal-in-time contributions, we split the PN scattering angle into local and nonlocal contributions. The total 4PN scattering angle is then given by 
\beq
\frac{\chi}{2} = \sum^{\infty}_{j=1} \frac{1}{j^n}\bigg(\chi_{\text{j,loc}}^{(n)}(\nu_{\infty}) + \chi_{\text{j,nloc}}^{(n)}(\nu_{\infty})\bigg)
\eeq
where $\nu_{\infty} = \sqrt{\gamma^2-1}$. For the nonlocal contribution we only include terms that are proportional to $\log \nu_\infty$. We also consider the 4PM scattering angle, including both conservative and radiative contributions \cite{4PM_paper1,4PM_paper2,4PM_paper3,4PM_paper4,4PM_paper5,4PM_paper6,4PM_paper7}. We split the components into conservative and radiative parts. We also include a $\mathcal{L}$-resummed PM scattering angle of the form introduced in \cite{B2B_paper1,strong_field_scattering}. The resummation procedure requires information about the boundary between scattering and non-scattering orbits. We note that the estimate for $J(E)$ on the boundary introduced in \cite{strong_field_scattering} does not work beyond a limited range of energies, so we instead use our own estimate for the separatrix from Eq.~\eqref{eq:separatrix_fit}. We also tested this resummation on the scattering information provided in \cite{B2B_paper3}, but found the 4PM resummed results to be significantly more accurate. For simplicity we will refer to the 4PN local contributions as ``4PN'' and the 4PM conservative contributions as ``4PM'' in all our figures. We also note that we found the radiative contributions in the PM expansion and the nonlocal contribution in the PN expansion for the scattering angle to be negligible, and we did not include them in our figures since the results were indistinguishable. Likewise, the resummed conservative 4PM and resummed full 4PM scattering angle showed negligible differences.

\begin{figure}
    \centering
    \includegraphics[trim={0 0 0 1cm},clip,trim={0 0 0 1cm},clip,width=0.5\textwidth]{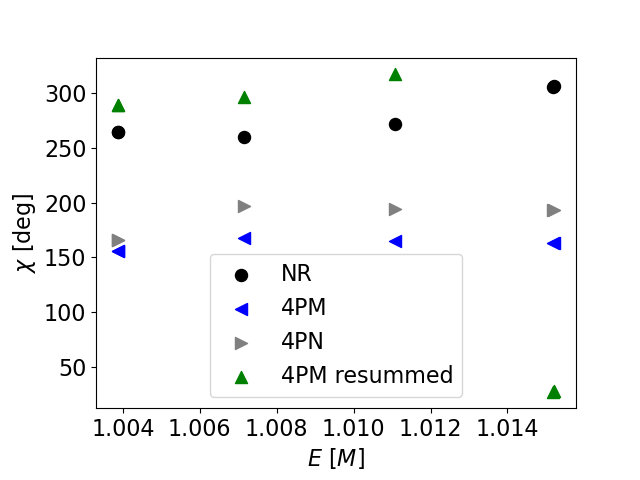}
    \caption{Scattering angle comparison for a selection of runs with $J = 1.07\,M^2$. The resummation technique is that of \cite{strong_field_scattering}.}
    \label{fig:scattering_comparison_J_1.069}
\end{figure}
    \begin{figure}
    \centering
    \includegraphics[trim={0 0 0 1cm},clip,width=0.5\textwidth]{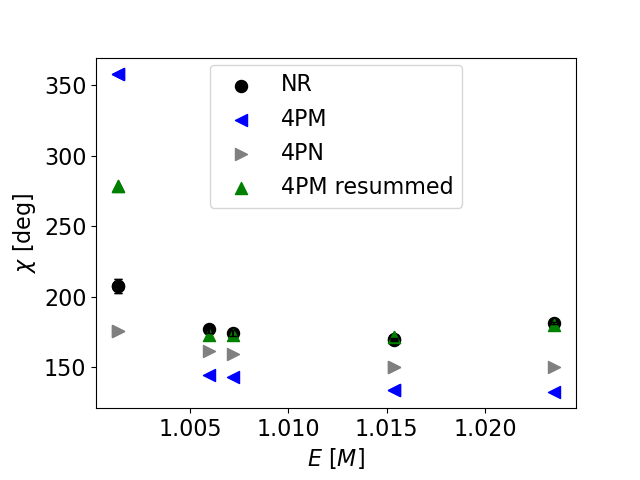}
    
    \caption{
    Same as Fig.~\ref{fig:scattering_comparison_J_1.069}, but with $J = 1.2\,M^2$.}
    \label{fig:scattering_comparison_J_1_dot_2}
\end{figure}
\begin{figure}
    \centering
    \includegraphics[trim={0 0 0 1cm},clip,width=0.5\textwidth]{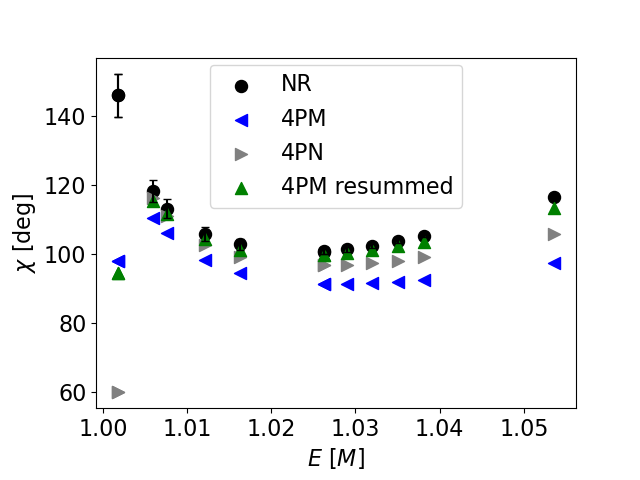}

    \caption{
    Same as Fig.~\ref{fig:scattering_comparison_J_1.069}, but with $J = 1.5\,M^2$.}
    \label{fig:scattering_comparison_J_1_dot_5}
\end{figure}
\begin{figure}
    \centering
    \includegraphics[trim={0 0 0 1cm},clip,width=0.5\textwidth]{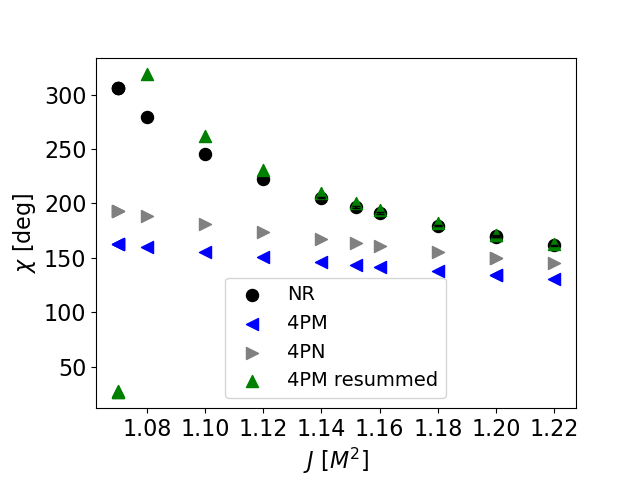}
    \caption{Scattering angle comparison for a selection of runs with $1.015\,M<E<1.016\,M$.}
    \label{fig:scattering_comparison_E_1_dot_015}
\end{figure}
\begin{figure}
    \centering
    \includegraphics[trim={0 0 0 1cm},clip,width=0.5\textwidth]{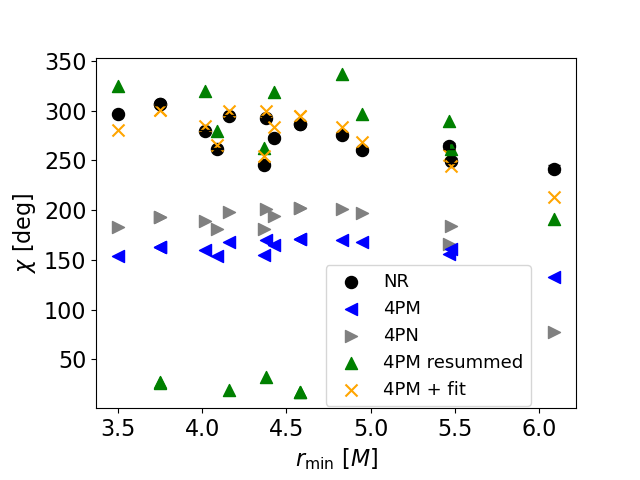}
    \caption{Scattering angle comparison for a selection of runs with $J<1.1\,M^2$. Here we also include a combination of 4PM and an NR calibrated fit (``4PM + fit'') using the form of Eq. \eqref{eq:4PM+fit}. We note each data point corresponds to a single unbound run, thus $r_\text{min}$ is the periastron distance of that specific run. }
    \label{fig:scattering_comparison_with_NR}
\end{figure}
Figs.~\ref{fig:scattering_comparison_J_1.069}--\ref{fig:scattering_comparison_with_NR} show comparisons with numerical results for the scattering angle for a selection of runs at various initial energies and angular momenta. Each data point now corresponds to a specific unbound run, and there is no binning process as in Figs.~\ref{fig:compare_B2B_arb_E}--\ref{fig:E_rad_compare_25M_eps6}. The resummed 4PM scattering angle shows a clear improvement over the PN and PM expanded scattering angle across much of the parameter space. However, for very low initial energies we see a clear divergence from the numerical data. Fig.~\ref{fig:scattering_comparison_J_1_dot_5} shows that as we lower the initial energy, the PN, PM and resummed PM scattering angles all diverge from the numerical result. Furthermore, Fig.~\ref{fig:scattering_comparison_E_1_dot_015} shows that at low initial angular momenta, there is still a clear difference between the resummed PM scattering angle and the numerical result. In the small $E$ and small $J$ section of the parameter space, significant improvements in the analytical sector are still needed, but the resummation procedure suggested in \cite{B2B_paper1,strong_field_scattering} shows excellent agreement with the numerical results over a large region of the parameter space.

Due to the difficulty in finding a simple NR fitting form that can uniformly improve the scattering angle across the parameter space, we instead opt to focus on runs with $J < 1.1\,M^2$. We do this because on the bound side, runs with $J < 1.1\,M^2$ are where we see the largest disagreements with the PN and PM analytical information, and generally correspond to our strongest interacting runs. It is therefore instructive to observe whether adding an NR fit can improve the results in this area of the parameter space. We adopt a functional form of the type
\beq
\chi = \chi_{\mathrm{4PM}} + \frac{aE^b}{j^6}
\label{eq:4PM+fit}
\eeq
and indeed, Fig.~\ref{fig:scattering_comparison_with_NR} shows agreement with the numerical results better than the resummed 4PM scattering angle.
\subsubsection{Periastron Advance}
\begin{figure}
    \centering
     \includegraphics[trim={0 0 0 1cm},clip,width=0.5\textwidth]{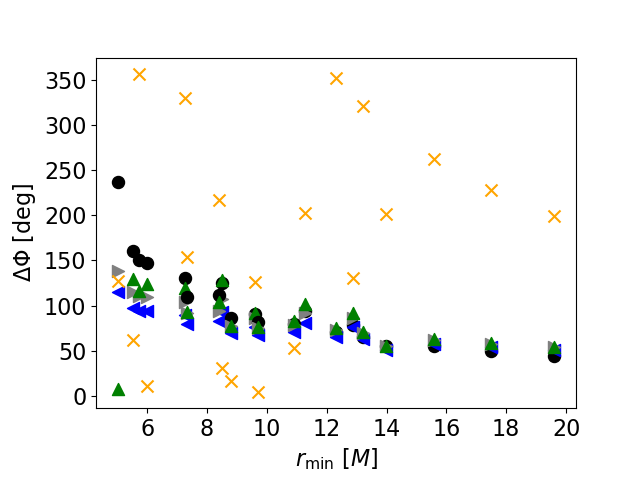}
    \caption{Comparisons of the periastron advance of the first orbit of NR bound systems to those obtained by applying the B2B relation for periastron advance, Eq. \eqref{eq:periastron_B2B_eq}, to various estimates for the scattering angle. Legend is the same as in Figs.~\ref{fig:scattering_comparison_with_NR} and \ref{fig:period_comparison}.}
    \label{fig:del_phi}
    \end{figure}

Similar to the relations for the radiated energy and angular momentum, the B2B developers introduced a non perturbative relation between the scattering angle and the periastron advance \cite{B2B_paper1,B2B_paper2}, 
\beq
\Delta \Phi (J,E) = \chi(J,E) + \chi (-J,E)\,.
\label{eq:periastron_B2B_eq}
\eeq
As can be seen from Fig.~\ref{fig:del_phi}, we see excellent agreement using unbound 4PM data with the numerical results for the large periastron runs. However, as the periastron distance becomes smaller, the analytical conservative prediction underestimates the periastron advance.

The NR fitting term, which was calibrated to unbound runs with $J < 1.1\,M^2$, does not show any agreement with the data. This clearly suggests that any generic form, even if it shows excellent agreement on the unbound side, does not transfer over smoothly to the the bound side. Interestingly, the resummed 4PM scattering angle does not show the wild fluctuations seen in the fitting term, but rather shows improvement over the 4PM result over certain regions of parameter space. Comparing the conservative 4PM to the full 4PM data including radiation reaction effects, we once again see no difference between the two. 

\subsubsection{Constructing the Radial Action}
Since no simple non-perturbative equation exists for the orbital period, we must first construct the radial action $S_r$,
\beq
    S_r(J,\varepsilon) = \mu\bigg( \text{sgn}(\hat{p}_\infty) \chi_j^{(1)}(\varepsilon) - j\bigg(1+\frac{2}{\pi} \sum_{n=1} \frac{\chi_j^{(2n)}(\varepsilon)}{(1-2n)j^{(2n)}}\bigg)\bigg)\,.
\eeq
From the radial action, we can take the appropriate derivative to obtain our corresponding dynamical invariant. The period is given by
\beq
T = \frac{2\pi}{\mu}\frac{\partial S_r(J,\varepsilon)}{\partial\varepsilon}
\label{eq:period_B2B_eq}
\eeq
and the periastron advance is given by 
\beq
\Phi = 2\pi + \Delta \Phi = -2\pi \frac{\partial S_r(J,\varepsilon)}{\partial \varepsilon}\,.
\eeq
Our general approach is to construct the radial action for unbound orbits, analytically continue to the bound regime, and differentiate with respect to energy to obtain the period. While a non-perturbative approach for construction of the radial action was proposed in \cite{B2B_paper1}, we opt to model the radial action in terms of the expanded scattering angle. The resummation acts as a modification of the $\chi^{(n)}_j$ terms, and an NR fitting term is simply the addition of a term in the summation.

\subsubsection{Period}
\begin{figure}
    \includegraphics[trim={0 0 0 1cm},clip,width=0.5\textwidth]{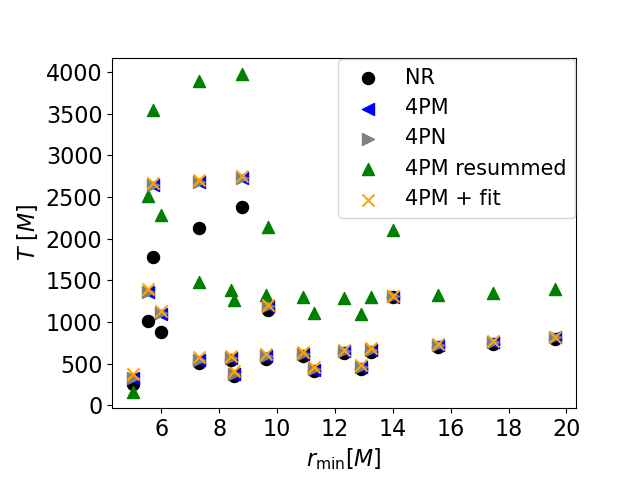}
    \caption{Comparisons of the period of the first orbit of NR bound orbits to those obtained by applying the B2B relation for period, Eq. \eqref{eq:period_B2B_eq}. on a radial action constructed from various scattering angle estimates.}
    \label{fig:period_comparison}
\end{figure}

 Fig.~\ref{fig:period_comparison} shows the result of applying the B2B relations using our various expressions for the scattering angle. Immediately we note that the 4PM, local, and NR fitting term perform almost identically, and further investigation reveals that the period is completely dominated by the 1PM scattering angle. For the fitting term we also tested a modification of the $\chi_j^{(1)}$ term, but again found the difference between the 4PM result to be negligible. The resummation on the other hand shows clear deviations from the numerical result and the 4PM result. This may be due to our approach of constructing the radial action.

\section{Summary and Discussion}
In this paper we tested the B2B relations presented in \cite{B2B_paper1,B2B_paper2,B2B_paper3} against numerical simulations of bound and unbound orbits of non-spinning equal mass black holes. We tested a wide range of eccentricities and tested relations for both radiative ($\Delta E$, $\Delta J$) and orbital observables ($T, \Delta \Phi$). We note that we were unable to produce unbound orbits with $J \lsim 1.05 \ M^2$, which results in limitations to the initial angular momentum of the bound orbits we can use to test the B2B map. We did not find strong evidence that the B2B relations hold in full GR. While we did find that, with the use of a specific fitting form, the B2B relations seemed to hold for bound orbits starting at 25 $M$, we do not believe this signifies any physical meaning, but rather is a result of our choice of fitting form. Likewise, for both the period and periastron advance, we were unable to find any evidence that the B2B relations \cite{B2B_paper1,B2B_paper2,B2B_paper3} continue to hold in full GR. 

In \cite{strong_field_scattering,strong_field_spinning_scattering}, similar to the approach employed here, the authors used an analytically informed and NR calibrated scattering function to extract the potential from scattering simulations. They did not attempt to transfer information between unbound and bound orbits, but only considered unbound systems. While in \cite{strong_field_scattering,strong_field_spinning_scattering} the higher order and radiative scattering corrections incorporated through an NR fitting term resulted in a more accurate calculation of the scattering potential, in this work we have shown that generic functional forms which fit the data very well on one side of the B2B correspondence do not transfer over robustly to the other side. 

We compared both PN and PM expanded scattering angle expressions against a large number of unbound orbits. We found that the resummation procedure suggested in \cite{B2B_paper1,strong_field_scattering} shows excellent agreement with numerical results over a large region of the parameter space. However, for unbound orbits starting with very low initial energy and angular momentum values, we see consistent disagreement with analytical expressions, and in some cases a strong divergence between numerical results and analytical estimates. 

Overall, our work indicates that for strongly radiating systems where higher-order radiative corrections become non-negligible, the B2B relations as currently proposed in the literature do not hold. This motivates further work to attempt to fully map non-perturbative corrections between bound and unbound systems. Our results therefore place clear limits on the applicability of the B2B relations for generating gravitational waveforms.

\begin{acknowledgments}
The authors thank Rafael Porto for helpful discussions. AK and STM were supported in part by NSF CAREER grant PHY-1945130, NASA grant 22-LPS22-0022 and 24-2024EPSCoR-0010. This research was made
possible by the NASA West Virginia Space Grant Consortium, Grant \# 80NSSC20M0055. This research was also supported in part by NSF grant PHY-1748958 to the Kavli Institute for Theoretical Physics (KITP). The authors acknowledge the computational resources provided by the WVU Research Computing Spruce Knob HPC cluster, which is funded in part by NSF EPS-1003907, and the Thorny Flat HPC cluster, which is funded in part by NSF OAC-1726534.

\end{acknowledgments}
\nocite{*}
\bibliography{citations}
\end{document}